\documentclass[sigplan,screen]{acmart}

\usepackage{multirow}

\usepackage[]{hyperref}
\texorpdfstring{}{}

\AtBeginDocument{%
  \providecommand\BibTeX{{%
    \normalfont B\kern-0.5em{\scshape i\kern-0.25em b}\kern-0.8em\TeX}}}

\setcopyright{acmcopyright}
\copyrightyear{2023}
\acmYear{2023}
\acmDOI{XXXXXXX.XXXXXXX}

\acmConference[PPoPP'23]{Make sure to enter the correct
  conference title from your rights confirmation emai}{February 25--March 1, 2023}{Montreal, Canada}
\acmPrice{15.00}
\acmISBN{978-1-4503-XXXX-X/18/06}

\begin{document}

\newcommand{\Name}{Fast-BNI}
\newcommand{\ZYW}[1]{\textcolor{red}{#1}}

\title{POSTER: Fast Parallel Exact Inference on Bayesian Networks}


\author{Jiantong Jiang\texorpdfstring{$^\dag$}{}, Zeyi Wen\texorpdfstring{$^{\ddag,\S}$}{}, Atif Mansoor\texorpdfstring{$^\dag$}{}, Ajmal Mian\texorpdfstring{$^\dag$}{}}
\affiliation{%
  \institution{\texorpdfstring{$^\dag$}{}The University of Western Australia\\
  \texorpdfstring{$^\ddag$}{}Hong Kong University of Science and Technology (Guangzhou); $^\S$Hong Kong University of Science and Technology}
  \country{}}
\email{jiantong.jiang@research.uwa.edu.au, wenzeyi@ust.hk, {atif.mansoor, ajmal.mian}@uwa.edu.au}

\renewcommand{\shortauthors}{J. Jiang et al.}


\begin{abstract}
Bayesian networks (BNs) are attractive, because they are graphical and interpretable machine learning models. 
However, exact inference on BNs is time-consuming, especially for complex problems. To improve the efficiency, we propose a fast BN exact inference solution named \Name{} on multi-core CPUs. 
\Name{} enhances the efficiency of exact inference through hybrid parallelism that tightly integrates coarse- and fine-grained parallelism. 
We also propose techniques to further simplify the bottleneck operations of BN exact inference. \Name{} source code is freely available at \url{https://github.com/jjiantong/FastBN}.

\end{abstract}


%

\maketitle

\section{Introduction}

Bayesian networks (BNs) are probabilistic graphical models. BNs use directed acyclic graphs (DAGs), which are often learned from data~\cite{jiang2022fast}, to represent random variables and their conditional dependencies.
\emph{Exact inference} on BNs is an essential task that calculates the conditional probability of certain \emph{query variables}, given some values of other variables called \emph{evidence} as knowledge to the BN.

Junction Tree (JT) is one of the most prominent BN exact inference algorithms. The key idea is to first convert a BN into a secondary structure called junction tree, where each node (called \emph{clique}) and edge (called \emph{separator}) in the tree contains a subset of variables and maintains a \emph{potential table} over the variables. It then passes \emph{messages} (i.e., functions over variables) along the tree structure and updates all the potential tables. 

However, exact inference on BNs is proven to be NP-hard. The complexity of JT increases dramatically with the clique sizes (i.e., the potential table sizes of the cliques), which hinders the use of BNs in complex problems.
There are two main types of approaches to accelerate JT on multi-core CPUs.
The first type uses coarse-grained inter-clique parallelism that parallelizes the message passing of different cliques~\cite{Kozlov1994}. However, inter-clique parallelism is load unbalanced, because the workloads for various cliques are highly different. Some approaches in this category utilize pointer jumping techniques, but introduce additional overhead caused by the rerooting or merging operation.
The second type of approaches is fine-grained intra-clique parallelism that parallelizes the potential table operations inside each clique~\cite{Xia2007}. Zheng~\cite{Zheng2013} accelerated JT on GPUs using the similar idea. However, this type of approaches has efficiency issues from the large parallelization overhead since the table operations are invoked frequently.
Moreover, inter-clique parallelism exhibits limited performance for the trees with a small number of cliques, and intra-clique parallelism has efficiency issues on the trees with many small cliques. Therefore, both can only be more efficient under certain junction tree structures.

\begin{table*}
    \centering
    \caption{Comparison of \Name{} with other implementations, and speedup of \Name{} over each compared implementation.}
    \begin{tabular}{|l||c|c|c||c|c|c|c|c|c|c|}
        \hline

        \multirow{3}{*}{BN} & 
        \multicolumn{3}{c||}{Sequential implementation} &
        \multicolumn{7}{c|}{Parallel implementation} \\
        \cline{2-11}
            & 
        \multicolumn{2}{c|}{Execution time (sec)} &
        Speedup &
        \multicolumn{4}{c|}{Execution time (sec)} &
        \multicolumn{3}{c|}{Speedup} \\
        \cline{2-11}
			& UnBBayes & \Name{}-seq & UnBBayes & Dir.
			& Prim. & Elem. & \Name{}-par 
			& Dir. & Prim. & Elem. \\
        \hline
        \hline
        Hailfinder & 28.3 & 4.0 & 7.1 & 3.0 & 3.2 & 4.0 & 2.5 & 1.2 & 1.3 & 1.6 \\
        Pathfinder & 319.2 & 68.9 & 4.6 & 40.5 & 23.6 & 27.8 & 11.1 & 3.6 & 2.1 & 2.5 \\
        Diabetes & 90961 & 6944 & 13.1 	& 3016 & 2311 & 3316 & 558.6 & 5.4 & 4.1 & 5.9 \\
        Pigs & 43714 & 3729 & 11.7 & 3353 & 1068 & 2380 & 221.7 & 15.1 & 4.8 & 10.7 \\
        Munin2 & 3054 & 2643 & 1.2 & 1951 & 934.7 & 1638 & 241.7 & 8.1 & 3.9 & 6.8 \\
        Munin4 & 258194 & 34198 & 7.6 	& 20364 & 10348 & 21398 & 3021 & 6.7 & 3.4 & 7.1 \\
        \hline
    \end{tabular}
    \label{tab_overall}
\end{table*}

\section{Our Fast-BNI Design}
\label{sec_para}

To address the efficiency issues of directly using only inter- or intra-clique parallelism, we propose \Name{}, a fast and parallel BN exact inference solution with hybrid inter- and intra-clique parallelism.

For the inter-clique parallelism, we first develop a breadth-first search based \emph{traversal method} to exploit parallelization opportunities across the tree structures. Our traversal method views all the cliques and separators as nodes of the tree and marks the layer where each of them is located. Secondly, we employ a \emph{root selection strategy} to construct a more balanced tree with the minimal number of layers to reduce the total number of parallelization invocations.

For the intra-clique parallelism, we identify and parallelize three dominant potential table operations, including potential table \emph{marginalization}, \emph{extension} and \emph{reduction}. The key step to the potential table operations is to find the index mappings between the original and the updated tables. Accordingly, the complexity of these operations depend on the potential table size, which increases dramatically with the number of random variables in the clique (or separator) and the number of states of the variables.
We develop the corresponding intra-clique primitives that parallelize the index mapping computations of different potential table entries.

We find some shortcomings of accelerating the JT algorithm using only one of the two granularities of parallelism, such as load unbalancing between threads, high parallelization overhead and structure-dependent parallel performance. To remedy the shortcomings, \Name{} utilizes a hybrid parallelism that closely integrates inter- and intra-clique parallelism by flattening the nested operations. At the beginning of each layer, all the potential table entries corresponding to this layer are packed to constitute one of the parallel tasks. The tasks are then distributed to the parallel threads to perform concurrently. To summarize, the proposed parallelism has three main advantages, including (i) workload balancing, (ii) smaller parallelization overhead and (iii) adaptability to various structures.

\section{Evaluation}
\label{sec_exp}

We conducted experiments on a Linux machine with two 26-core 2GHz Intel Xeon Platinum 8167M CPUs and 768GB main memory.
Our algorithm was tested on six real-world BNs\footnote{https://www.bnlearn.com/bnrepository/} with different sizes, where the last four are considered as large-scale BNs. 
We randomly generated 2,000 test cases from each network, each with 20\% of the observed variables.  

Table~\ref{tab_overall} shows the execution time comparison of sequential and parallel implementations of \Name{} with the existing implementations. Specifically, we compared the sequential version of \Name{} (i.e., \Name{}-seq) with the JT implementation in the UnBBayes library\footnote{UnBBayes: \url{https://sourceforge.net/projects/unbbayes/}.}~\cite{Carvalho2010}; we also compared \Name{} parallel version (i.e., \Name{}-par) with parallel implementations~\cite{Kozlov1994, Xia2007, Zheng2013}. Implementation~\cite{Kozlov1994} (denoted by \emph{Direct}, \emph{Dir.}) uses a \emph{direct} coarse-grained parallelism; implementation~\cite{Xia2007} (denoted by \emph{Primitive}, \emph{Prim.}) proposes four fine-grained node-level \emph{primitives} for JT; implementation~\cite{Zheng2013} (denoted by \emph{Element}, \emph{Elem.}) utilizes fine-grained \emph{element}-wise parallelism.
For comparing the parallel implementations, we varied the number of OpenMP threads $t$ from 1 to 32 and chose the one with the shortest execution time.

As can be seen from the ``Speedup'' columns of Table~\ref{tab_overall}, the sequential implementation of \Name{} can be 1.2 to 13.1 times faster than UnBBayes.
When comparing the parallel implementations, \Name{}-par can run 1.2 to 15.2 times faster than the counterparts. 
It is worth noting that \Name{} has more advantages over existing implementations on larger networks. 
For some small-scale networks, the speedups of \Name{} to other parallel implementations are relatively small, because they require short execution time for Bayesian inference (e.g., less than 4 seconds for \emph{Hailfinder}) and the parallelization overhead of the small-scale networks accounts for a large proportion.
Another observation is that \Name{} always achieves its shortest execution time when $t = 32$ on large BNs.
\emph{Munin4} is a very large BN with more than 1,000 nodes and edges. The experiment on \emph{Munin4} is the task that takes the longest time to complete. This task ran almost three days using UnBBayes, and spent 3 to 6 hours using the existing parallel implementations, while the execution time is significantly reduced to less than one hour using the proposed \Name{}.

\begin{acks}
Professor Ajmal Mian is the recipient of an Australian Research Council Future Fellowship Award (project number FT210100268) funded by the Australian Government. Zeyi Wen is the corresponding author.
\end{acks}

\balance

\bibliographystyle{ACM-Reference-Format}
\bibliography{all}

\end{document}